\documentclass[12pt]{article}
\usepackage{amsfonts}
\usepackage{amsmath}
\usepackage{bm}
\usepackage{graphicx}

\newcommand{\eqdef}{\stackrel{\rm def}{=}}
\newcommand{\sfrac}[2]{{\textstyle \frac{#1}{#2}}}

\begin{document}

\baselineskip=20pt

\newfont{\elevenmib}{cmmib10 scaled\magstep1}
\newcommand{\preprint}{
    \begin{flushleft}
      \elevenmib Yukawa\, Institute\, Kyoto\\
    \end{flushleft}\vspace{-1.3cm}
    \begin{flushright}\normalsize  \sf
      DPSU-04-4\\
      YITP-04-60\\
      {\tt hep-th/0410109} \\ October 2004
    \end{flushright}}
\newcommand{\Title}[1]{{\baselineskip=26pt
    \begin{center} \Large \bf #1 \\ \ \\ \end{center}}}
\newcommand{\Author}{\begin{center}
    \large \bf Satoru Odake${}^a$ and Ryu Sasaki${}^b$ \end{center}}
\newcommand{\Address}{\begin{center}
      $^a$ Department of Physics, Shinshu University,\\
      Matsumoto 390-8621, Japan\\
      ${}^b$ Yukawa Institute for Theoretical Physics,\\
      Kyoto University, Kyoto 606-8502, Japan
    \end{center}}
\newcommand{\Accepted}[1]{\begin{center}
    {\large \sf #1}\\ \vspace{1mm}{\small \sf Accepted for Publication}
    \end{center}}

\preprint
\thispagestyle{empty}
\bigskip\bigskip\bigskip

\Title{Equilibrium Positions, Shape Invariance and Askey-Wilson Polynomials}
\Author

\Address
\vspace{1cm}

\begin{abstract}
We show that the equilibrium positions of the Ruijsenaars-Schneider-van Diejen
systems with the trigonometric potential are given by the zeros of the
Askey-Wilson polynomials with five parameters.
The corresponding single particle quantum version, which is a typical example
of ``discrete" quantum mechanical systems with a $q$-shift type kinetic term,
is shape invariant and the eigenfunctions are the Askey-Wilson polynomials.
This is an extension of our previous study \cite{os2,os4},
which established the ``discrete analogue" of the well-known fact;
The equilibrium positions of the Calogero systems are described by the
Hermite and Laguerre polynomials, whereas the corresponding single particle
quantum versions are shape invariant and the eigenfunctions are the Hermite 
and Laguerre polynomials.
\end{abstract}

\newpage
\section{Introduction}
\label{intro}

The Calogero-Sutherland systems \cite{Cal-Sut} and their integrable
deformation called the Ruijsenaars-Schneider-van Diejen systems
\cite{RS,vD} have many attractive features at both classical
and quantum mechanical levels.
In our recent paper \cite{rs,os2}, the equilibrium positions of the
classical Ruijsenaars-Schneider-van Diejen systems were studied.
The equilibrium positions of the Calogero-Sutherland systems are
described by the zeros of the classical orthogonal polynomials,
the Hermite, Laguerre, Chebyshev, Legendre, Gegenbauer and Jacobi
polynomials \cite{calmat,sti,os1}.
Since the Ruijsenaars-Schneider-van Diejen systems are deformation of
the Calogero-Sutherland systems, it is expected that
the equilibrium positions of the Ruijsenaars-Schneider-van Diejen
systems are described by some deformation of these classical orthogonal
polynomials. This is indeed the case and we obtained the deformed
Hermite, Laguerre and Jacobi polynomials \cite{os2}. These deformed
orthogonal polynomials fit in the Askey-scheme of the hypergeometric
orthogonal polynomials \cite{And-Ask-Roy,koeswart};
(\romannumeral1) rational potential cases:
One and two parameter deformation of the Hermite polynomials are
a special case of the Meixner-Pollaczek polynomial and a special case of
the continuous Hahn polynomial,
and two and three parameter deformation of the Laguerre polynomials are
the continuous dual Hahn polynomial and the Wilson polynomial,
(\romannumeral2) trigonometric potential cases: Several one parameter
deformation of the Jacobi polynomials are special cases of the
Askey-Wilson polynomial.
The Askey-Wilson polynomial has five parameters \cite{AW},
but the deformed Jacobi polynomials obtained in \cite{os2} have only three
parameters.
A natural question arises;
Find (integrable) multi-particle systems whose equilibrium positions are
described by the Askey-Wilson polynomials with five parameters.

Shape invariance is an important ingredient of many exactly solvable
quantum mechanics \cite{genden,crum,spivinzhed}.
In another recent paper of ours \cite{os4}, the shape invariance of
``discrete" quantum mechanical single particle systems, whose kinetic
term causes a shift of the coordinate in the imaginary direction,
are discussed.
The eigenfunctions of these shape invariant systems are
a special case of the Meixner-Pollaczek polynomial, 
a special case of the continuous Hahn polynomial,
the continuous dual Hahn polynomial and the Wilson polynomial.
These polynomials have all appeared in the above discussion about the
equilibrium positions, in which we have one more polynomial,
the Askey-Wilson polynomial.
This gives the second question;
Are the quantum mechanical single particle systems, whose eigenstates are
the Askey-Wilson polynomial, shape invariant or not ?

We will answer the above two questions in this paper.
The answer to the first question is found by the same method given in
\cite{os2}, \emph{i.e.\/} numerical analysis, functional equation and
three-term recurrence.
The second question is answered affirmatively by using the properties of
the Askey-Wilson polynomials and similar discussion in \cite{os4} with
replacement of a shift operator by a $q$-shift operator.

This article is organized as follows.
In section 2, the classical equilibria of the
Ruijsenaars-Schneider-van Diejen system are studied. For a suitable
choice of the elementary potential functions, the equilibrium positions are
given by the zeros of the Askey-Wilson polynomials with five parameters.
In section 3 we discuss the shape invariance of ``discrete" quantum
mechanical single particle systems with a $q$-shift type kinetic term.
After discussing general theory, we present an explicit example of such
shape invariant systems, in which eigenstates are described by the
Askey-Wilson polynomials.
The final section is for a summary and comments.

\section{Multi-particle Systems: Equilibrium Positions}
\label{multiparticle}

Let us consider the equilibrium positions of the classical
Ruijsenaars-Schneider-van~Diejen
systems with the trigonometric potential \cite{RS,vD},
which are integrable deformation of the celebrated Calogero-Sutherland
systems of exactly solvable multi-particle quantum mechanics.
Its classical Hamiltonian corresponding to the BC root system
is the following \cite{vD}:
\begin{align}
  H(p,q)&=\sum_{j=1}^n\left( \cosh p_j\,\sqrt{V_j(q)\,{V}_j^*(q)}
  -\frac12\Bigl(V_j(q)+{V}_j^*(q)\Bigr)\right),
  \label{H(p,q)}\\
  V_j(q)&=w(q_j)\prod_{k=1\atop k\neq j}^n v(q_j-q_k)\,v(q_j+q_k)
  \qquad(j=1,\ldots,n),\\
  v(x)&=\frac{\sin(x-ig_0)}{\sin x}\,,\\
  w(x)&=\frac{\sin(x-ig_1)}{\sin x}\,\frac{\sin(x-ig_2)}{\sin x}\,
        \frac{\cos(x-ig_3)}{\cos x}\,\frac{\cos(x-ig_4)}{\cos x}\,,
\end{align}
where $q={}^t(q_1,\cdots,q_n)$ and $p={}^t(p_1,\cdots,p_n)$ are
the coordinates and conjugate momenta,
and $g_j$ ($j=0,\ldots,4$) are the real positive coupling constants.
The potentials $V_j$ and $V_j^*$ are  complex conjugate of each other.
Our convention of a complex conjugate function is the following:
for an arbitrary function $f(x)=\sum_na_nx^n$ ($a_n\in\mathbb{C}$),
we define $f^*(x)=\sum_na_n^*x^n$. Here $c^*$ is the complex conjugation
of a number $c\in\mathbb{C}$.
Note that $f^*(x)$ is not the complex conjugation of $f(x)$,
$(f(x))^*=f^*(x^*)$.
This is relevant for considering complex variables in section
\ref{singleparticle}.
The equilibrium positions $p=0$, $q=\bar{q}$ are determined by the condition
\cite{rs}
\begin{equation}
  V_j(\bar{q})=V_j^*(\bar{q})>0\qquad(j=1,2,\ldots,n).
  \label{equiveq}
\end{equation}
This equation {\em without inequality\/} is rewritten in the Bethe
ansatz like equation
\begin{equation}
  \prod_{k=1\atop k\neq j}^n
  \frac{v(\bar{q}_j-\bar{q}_k)v(\bar{q}_j+\bar{q}_k)}
  {v^*(\bar{q}_j-\bar{q}_k)v^*(\bar{q}_j+\bar{q}_k)}
  =\frac{w^*(\bar{q}_j)}{w(\bar{q}_j)}\,.
  \label{equiveqBA}
\end{equation}
Note that $\bar{q}_j=0,\frac{\pi}{2}$ is excluded in (\ref{equiveq}) but
allowed in (\ref{equiveqBA}).

By the same method given in \cite{os2} (numerical analysis, functional
equation and three-term recurrence), we can show that the equilibrium positions
$\{\bar{q}_j\}$ are given by the zeros of the Askey-Wilson polynomial
\cite{AW} (we follow the notation of Koekoek and Swarttouw \cite{koeswart})
$p_n(x;a,b,c,d|q)\propto\prod_{j=1}^n(x-\cos 2\bar{q}_j)$ with the
following parameters,\footnote{
While preparing our manuscript, we became aware of a paper by van Diejen
\cite{vD04}, in which the same result was presented with a more elegant
proof.
}
\begin{equation}
  q=e^{-2g_0},\quad (a,b,c,d)=(e^{-2g_1},e^{-2g_2},-e^{-2g_3},-e^{-2g_4}).
\end{equation}
The outline of derivation is as follows. The functional equation for
$f(x)=\prod_{j=1}^n(x-\cos 2\bar{q}_j)$ is the same as
eq.(A.4)($\epsilon=-1$) in \cite{os2} with\footnote{ 
This form of $g(x)$ is obtained by using some empirical knowledge based
on numerical analysis.
}
\begin{align}
  h(x)&=\Bigl(\sfrac{q^{-1}-q}{2}x-i(1+\sfrac{q^{-1}+q}{2})\sqrt{1-x^2}\Bigr)
  \nonumber\\
  &\quad\times
  \Bigl(\sfrac{a^{-1}-a}{2}(1+x)+i(1+\sfrac{a^{-1}+a}{2})\sqrt{1-x^2}\Bigr)
  \Bigl(\sfrac{b^{-1}-b}{2}(1+x)+i(1+\sfrac{b^{-1}+b}{2})\sqrt{1-x^2}\Bigr)
  \nonumber\\
  &\quad\times
  \Bigl(\sfrac{c-c^{-1}}{2}(1-x)-i(1-\sfrac{c+c^{-1}}{2})\sqrt{1-x^2}\Bigr)
  \Bigl(\sfrac{d-d^{-1}}{2}(1-x)-i(1-\sfrac{d+d^{-1}}{2})\sqrt{1-x^2}\Bigr),\\
  g_n(x)&=\frac{(1+a)(1+b)(1-c)(1-d)(1+q)(1-x^2)}{8abcdq^{n+1}}\nonumber\\
  &\quad\times\Bigl(
  4q(1+abcdq^{2n-1})x^2 
  -2q^n\bigl(abc+abd+acd+bcd+(a+b+c+d)q\bigr)x\nonumber\\
  &\quad\qquad
  -(1+q)\bigl(1+q-q^{n+1}-(ab+ac+ad+bc+bd+cd)q^n \nonumber\\
  &\quad\qquad\qquad\qquad\quad
  -q^{n-1}(1-q^n-q^{n+1})abcd\bigr)\Bigr),
\end{align}
and $\sqrt{\delta}=\frac{1-q}{1+q}$.
The three-term recurrence $f_{n+1}(x)=(x-a_n)f_n(x)-b_nf_{n-1}(x)$ 
for the monic Askey-Wilson polynomial can be found in the literature, 
for example eq.(3.1.5) in \cite{koeswart}.
Then eq.(A.7) in \cite{os2} holds because 
the functions $X_n(x),Y_n(x)$ of (A.13),(A.14) in \cite{os2} vanish
for these $a_n,b_n$ and $g_n(x)$. 
Therefore we obtain Proposition A.3, A.4 in \cite{os2}.

\noindent
Remark 1: The models studied in \cite{os2} are special cases of this
model. For example,
\begin{eqnarray*}
  \mbox{$B$ eq.(2.75) in \cite{os2}}&&
  (g_0,g_1,g_2,g_3,g_4)=(g_L,\sfrac12 g_S,\sfrac12 g_S,0,0),\\
  \mbox{$C'$ eq.(2.78) in \cite{os2}}&&
  (g_0,g_1,g_2,g_3,g_4)=(g_S,\sfrac12 g_L,\sfrac12 g_L,\sfrac12 g_L,
  \sfrac12 g_L),\\
  \mbox{$B'C$ eq.(2.79) in \cite{os2}}&&
  (g_0,g_1,g_2,g_3,g_4)=(g_M,g_S,g_L,g_L,0),
\end{eqnarray*}
due to the following trigonometric formulas,
\begin{eqnarray*}
  &&\frac{\sin(x-ig)}{\sin x}=\cosh g\,(1-i\tanh g\cot x),\quad
  \frac{\cos(x-ig)}{\cos x}=\cosh g\,(1+i\tanh g\tan x),\\
  &&\cosh 2g(1-i\tanh 2g\cot 2x)=\frac{\sin(2x-i2g)}{\sin 2x}
  =\frac{\sin(x-ig)}{\sin x}\,\frac{\cos(x-ig)}{\cos x}.
\end{eqnarray*}

\noindent
Remark 2: Ismail et.al. studied the $q$-Strum Liouville problems and the
Bethe ansatz equation of the XXZ model \cite{ismail}.
A special case of their results states that the zeros of Askey-Wilson
polynomial $p_n(x;a,b,c,d|q)\propto\prod_{j=1}^n(x-\cos 2\bar{q}_j)$ with
the parameters
\begin{equation}
  q=e^{-2g_0},\quad (a,b,c,d)=(e^{-2g_1},e^{-2g_2},e^{-2g_3},e^{-2g_4}),
\end{equation}
satisfies the Bethe ansatz like equation (\ref{equiveqBA}) with
($g_j>0$)
\begin{equation}
  v(x)=\frac{\sin(x-ig_0)}{\sin x}\,,\quad
  w(x)=\frac{\sin(x-ig_1)}{\sin x}\,\frac{\sin(x-ig_2)}{\sin x}\,
        \frac{\sin(x-ig_3)}{\sin x}\,\frac{\sin(x-ig_4)}{\sin x}\,.
  \label{ismailvw}
\end{equation}
However this does not mean that $\{\bar{q}_j\}$ are equilibrium
positions of the system (\ref{H(p,q)}) with (\ref{ismailvw}),
because $V_j(\bar{q})$ is not positive in this case.
Moreover the $\cot^4x$ term in $w(x)$  (\ref{ismailvw}) appears
too singular to give a satisfactory quantum  Hamiltonian.

\section{Single Particle Systems: Shape Invariance}
\label{singleparticle}

Next let us consider the shape invariance of the ``discrete"
quantum mechanical single particle systems with a $q$-shift type kinetic term.
The argument is parallel to that given in \cite{os4}, in which
discrete quantum systems with a shift-type kinetic term are discussed.

In this section we use variables $\theta$, $x$ and $z$, which are
related as
\begin{equation}
  0\leq\theta\leq\pi,\quad x=\cos\theta,\quad z=e^{i\theta}.
\end{equation}
The dynamical variable is $2\theta$ and the inner product is
$(f(\theta),g(\theta))=\int_0^{\pi}d\theta f(\theta)^*g(\theta)$.
We denote $D=D_z\eqdef z\frac{d}{dz}$.
Then $q^D$ is a $q$-shift operator, $q^Df(z)=f(qz)$.
Note that
\begin{equation}
  \int_0^{\pi}d\theta=\int_{-1}^1\frac{dx}{\sqrt{1-x^2}},\quad
  -i\frac{d}{d\theta}=z\frac{d}{dz}=D,\quad
  f(z)^*=f^*(z^{-1}).
\end{equation}

For a real constant $q$ ($0<q<1$) and a function $V(z)=V(z;\bm{\lambda},q)$
with a set of real parameters $\bm{\lambda}$,
let us consider the following Hamiltonian $H=H(z;\bm{\lambda},q)$,
\begin{equation}
  H\eqdef\sfrac12\sqrt{V(z)}\,q^{D}\!\sqrt{V^*(z^{-1})}
  +\sfrac12\sqrt{V^*(z^{-1})}\,q^{-D}\!\sqrt{V(z)}
  -\sfrac12(V(z)+V^*(z^{-1})).
  \label{H}
\end{equation}
The eigenvalue equation reads
\begin{equation}
  H\phi_n=\mathcal{E}_n\phi_n,
\end{equation}
with eigenfunctions $\phi_n(z)=\phi_n(z;\bm{\lambda},q)$
and eigenvalues $\mathcal{E}_n=\mathcal{E}_n(\bm{\lambda},q)$
($n=0,1,\ldots$)
(we assume non-degeneracy $\mathcal{E}_0<\mathcal{E}_1<\cdots$).
The kinetic term causes a $q$-shift in the variable $z$.
This Hamiltonian is factorized and consequently positive semi-definite:
\begin{align}
  H&=A^{\dagger}A,
  \label{HAdA}\\
  A&=A(z;\bm{\lambda},q)
  \eqdef\frac{1}{\sqrt{2}}\Bigl(q^{\frac{D}{2}}\sqrt{V^*(z^{-1})}
  -q^{-\frac{D}{2}}\sqrt{V(z)}\Bigr),\\
  A^{\dagger}&=A(z;\bm{\lambda},q)^{\dagger}
  \eqdef\frac{1}{\sqrt{2}}\Bigl(\sqrt{V(z)}\,q^{\frac{D}{2}}
  -\sqrt{V^*(z^{-1})}\,q^{-\frac{D}{2}}\Bigr),
\end{align}
where $\dagger$ denotes the hermitian conjugation with respect to the above
inner product.
The ground state $\phi_0$ is the function annihilated by $A$:
\begin{equation}
  A\phi_0=0\quad(\Rightarrow H\phi_0=0,\ {\cal E}_0=0).
  \label{ground}
\end{equation}
Explicitly this equation reads
\begin{equation}
  \sqrt{V^*(q^{-\frac12}z^{-1})}\,\phi_0(q^{\frac12}z)
  =\sqrt{V(q^{-\frac12}z)}\,\phi_0(q^{-\frac12}z).
\end{equation}
The other eigenfunctions can be obtained in the form
\begin{equation}
  \phi_n(z)\propto P_n(z) \phi_0(z),
  \label{excited}
\end{equation}
where $P_n(z)=P_n^{(\bm{\lambda},q)}(z)$ is a Laurent polynomial in $z$
(for the explicit example below, it is the Askey-Wilson polynomial in $x$).
This $P_n(z)$ satisfies
\begin{equation}
  \tilde{H}P_n=\mathcal{E}_n P_n.
  \label{tHP=EP}
\end{equation}
Here $\tilde{H}=\tilde{H}(z;\bm{\lambda},q)$ is a similarity transformed
Hamiltonian in terms of the ground state wavefunction $\phi_0$:
\begin{equation}
  \tilde{H}\eqdef\phi_0^{-1}\circ H\circ\phi_0
  =\sfrac12V(z)\,q^{D}+\sfrac12V^*(z^{-1})\,q^{-D}
  -\sfrac12(V(z)+V^*(z^{-1})).
  \label{tildeH}
\end{equation}
Corresponding to the factorization of $H$ (\ref{HAdA}), $\tilde{H}$ is
also factorized:
\begin{align}
  \tilde{H}&=BC,
  \label{tH=BC}\\
  C&=C(z,q)=\sfrac{1}{2}(q^{\frac{D}{2}}-q^{-\frac{D}{2}}),\\
  B&=B(z;\bm{\lambda},q)=V(z)\,q^{\frac{D}{2}}
  -V^*(z^{-1})\,q^{-\frac{D}{2}}.
\end{align}

Let us define a new set of wavefunctions
$\phi_{1,n}(z)=\phi_{1,n}(z;\bm{\lambda},q)$,
\begin{equation}
  \phi_{1,n}\eqdef A\phi_n\quad (n=1,2,\ldots).
\end{equation}
As a consequence of the factorization, they form eigenfunctions of
a new Hamiltonian $H_1=H_1(z;\bm{\lambda},q)$,
\begin{equation}
  H_1=AA^\dagger
\end{equation}
with the same eigenvalues $\{\mathcal{E}_n\}$:
\begin{equation}
  H_1\phi_{1,n}=AA^\dagger A\phi_n=A\mathcal{E}_n\phi_n
  =\mathcal{E}_n\phi_{1,n}\quad (n=1,2,\ldots).
\end{equation}
To consider the shape invariance of $H$, we try to find the operators $A_1$,
$A_1^{\dagger}$ and a real constant $\mathcal{E}_1$ satisfying
\begin{align}
  H_1&=AA^{\dagger}=A_1^{\dagger}A_1+{\cal E}_1,
  \label{H1}\\
  A_1&=A_1(z;\bm{\lambda},q)
  \eqdef\frac{1}{\sqrt{2}}\Bigl(q^{\frac{D}{2}}\sqrt{V_1^*(z^{-1})}
  -q^{-\frac{D}{2}}\sqrt{V_1(z)}\Bigr),\\
  A_1^{\dagger}&=A_1(x;\bm{\lambda},q)^{\dagger}
  \eqdef\frac{1}{\sqrt{2}}\Bigl(\sqrt{V_1(z)}\,q^{\frac{D}{2}}
  -\sqrt{V_1^*(z^{-1})}\,q^{-\frac{D}{2}}\Bigr).
\end{align}
In other words, given $V(z)=V(z;\bm{\lambda},q)$, find
a new potential $V_1(z)=V_1(z;\bm{\lambda},q)$ satisfying
\begin{gather}
  V_1(z)V_1^*(q^{-1}z^{-1})=V(q^{\frac12}z)V^*(q^{-\frac12}z^{-1}),
  \label{V1eq1}\\
  V_1(z)+V_1^*(z^{-1})=V(q^{-\frac12}z)+V^*(q^{-\frac12}z^{-1})+2{\cal E}_1.
  \label{V1eq2}
\end{gather}
If $V_1$ has the same functional form as $V$ with another set of
parameters $\bm{\lambda}'$ (e.g, $q$-shifted $\bm{\lambda}$),
\begin{equation}
  V_1(z;\bm{\lambda},q)\propto V(z;\bm{\lambda}',q),
\end{equation}
then it is shape invariant.
Suppose $V_1$ has the form
\begin{equation}
  V_1(z)=V(q^{\frac12}z)g(z),
  \label{V1=Vg}
\end{equation}
with an as yet unspecified function $g(z)$,
 the above conditions (\ref{V1eq1}), (\ref{V1eq2}) get slightly
simplified:
\begin{gather}
  g(z)g^*(q^{-1}z^{-1})=1,
  \label{V1eq3}\\
  V(q^{\frac12}z)g(z)+V^*(q^{\frac12}z^{-1})g^*(z^{-1})
  =V(q^{-\frac12}z)+V^*(q^{-\frac12}z^{-1})+2{\cal E}_1.
  \label{V1eq4}
\end{gather}
If the desired $V_1$ is found, we can construct $H_2,H_3,\cdots$ by
repeating the same step.
We illustrate this procedure by taking the Askey-Wilson polynomial as an
example.

Let us take $V$ as
\begin{equation}
  V(z)=V(z;\bm{\lambda},q)=\frac{(1-az)(1-bz)(1-cz)(1-dz)}{(1-z^2)(1-qz^2)},
\end{equation}
where $\bm{\lambda}=(a,b,c,d)$. For simplicity we assume $-1<a,b,c,d<1$. 
Note that $V^*(z)=V(z)$.
The ground state (\ref{ground}) is given by \cite{koeswart}
\begin{align}
  \phi_0(z)=\phi_0(z;\bm{\lambda},q)\propto
  &\sqrt{w(z;\bm{\lambda},q)}\eqdef\left|
  \frac{(z^2;q)_{\infty}}{(az,bz,cz,dz;q)_{\infty}}\right|\\[8pt]
  &=\sqrt{\frac{(z^2,z^{-2};q)_{\infty}}
  {(az,az^{-1},bz,bz^{-1},cz,cz^{-1},dz,dz^{-1};q)_{\infty}}}\,,
\end{align}
where $(a_1,\cdots,a_m;q)_{\infty}=\prod_{j=1}^m
\prod_{n=0}^{\infty}(1-a_jq^n)$.
Excited states have the form (\ref{excited})
$\phi_n(z)\propto P_n(z) \phi_0(z)$, and (\ref{tHP=EP}) implies
that $P_n(z)$ is proportional to the Askey-Wilson polynomial
\cite{koeswart},
\begin{gather}
  P_n(z)=P_n^{(\bm{\lambda},q)}(z)\propto p_n(x;a,b,c,d;q),\\
  \mathcal{E}_n=\mathcal{E}_n(\bm{\lambda},q)
  =\sfrac12q^{-n}(1-q^n)(1-abcdq^{n-1}),
\end{gather}
which is an orthogonal polynomial,
\begin{equation}
 \int_{-1}^1\frac{dx}{\sqrt{1-x^2}}\,w(z;\bm{\lambda},q)\,
 p_n(x;a,b,c,d|q)\,p_m(x;a,b,c,d|q)\propto\delta_{nm},
\end{equation}
namely $(\phi_n,\phi_m)\propto\delta_{nm}$.
By denoting $p_n(x;a,b,c,d|q)=P_n(z;\bm{\lambda},q)$,
the factorization (\ref{tH=BC}) gives the forward and backward shift
relations ((3.1.8) and (3.1.10) in \cite{koeswart}):
\begin{gather}
  C(z;q)P_n(z;\bm{\lambda},q)=-{\cal E}_n(\bm{\lambda},q)q^{\frac{n}{2}}
  (z-z^{-1})P_{n-1}(z;q^{\frac12}\bm{\lambda},q),\\
  -B(z;\bm{\lambda},q)\ q^{\frac{n}{2}}(z-z^{-1})
  P_{n-1}(z;q^{\frac12}\bm{\lambda},q)
  =P_n(z;\bm{\lambda},q).
\end{gather}

It is easy to check that $V_1$ in the form (\ref{V1=Vg}) with
\begin{equation}
  g(z)=q^{-1}\frac{1-q^2z^2}{1-z^2}
  \label{g}
\end{equation}
satisfies (\ref{V1eq3}) and (\ref{V1eq4}), and it becomes
\begin{equation}
  V_1(z;\bm{\lambda},q)=V(q^{\frac12}z)g(z)
  =q^{-1}V(z;q^{\frac12}\bm{\lambda},q),
  \label{V1=qV}
\end{equation}
and $\mathcal{E}_1$ is
\begin{equation}
  \mathcal{E}_1(\bm{\lambda},q)=\frac12q^{-1}(1-q)(1-abcd).
  \label{E1}
\end{equation}
Therefore we have shape invariance, (\ref{V1=qV}) and
\begin{align}
  A_1(z;\bm{\lambda},q)&=q^{-\frac12}A(z;q^{\frac12}\bm{\lambda},q),\\
  H_1(z;\bm{\lambda},q)&=q^{-1}H(z;q^{\frac12}\bm{\lambda},q)
  +\mathcal{E}_1(\bm{\lambda},q).
\end{align}
We write down important formulas once again:
\begin{gather}
  A(z;\bm{\lambda},q)\phi_0(z;\bm{\lambda},q)=0,\\
  A(z;\bm{\lambda},q)A(z;\bm{\lambda},q)^{\dagger}
  =q^{-1}A(z;q^{\frac12}\bm{\lambda},q)^{\dagger}
  A(z;q^{\frac12}\bm{\lambda},q)+\mathcal{E}_1(\bm{\lambda},q).
  \label{AAd=AdA+E}
\end{gather}

Starting from $V_0=V$, $H_0=H$, $\phi_{0,n}=\phi_n$, let us define
$V_s$, $H_s$, $\phi_{s,n}$ ($n\geq s\geq 0$) step by step:
\begin{align}
  V_{s+1}(z;\bm{\lambda},q)&\eqdef
  q^{-1}V_s(z;q^{\frac12}\bm{\lambda},q),\\
  H_{s+1}(z;\bm{\lambda},q)&\eqdef
  A_s(z;\bm{\lambda},q)A_s(z;\bm{\lambda},q)^{\dagger}
  +\mathcal{E}_s(\bm{\lambda},q),\\
  \phi_{s+1,n}(z;\bm{\lambda},q)&\eqdef
  A_s(z;\bm{\lambda},q)\phi_{s,n}(z;\bm{\lambda},q).
  \label{phi=Aphi}
\end{align}
Here $A_s$ and $A_s^{\dagger}$ are defined by
\begin{align}
  A_s(z;\bm{\lambda},q)&\eqdef\frac{1}{\sqrt{2}}\Bigl(
  q^{\frac{D}{2}}\sqrt{V_s^*(z^{-1};\bm{\lambda},q)}
  -q^{-\frac{D}{2}}\sqrt{V_s(z;\bm{\lambda},q)}\Bigr),\\
  A_s(z;\bm{\lambda},q)^{\dagger}&\eqdef\frac{1}{\sqrt{2}}\Bigl(
  \sqrt{V_s(z;\bm{\lambda},q)}\,q^{\frac{D}{2}}
  -\sqrt{V_s^*(z^{-1};\bm{\lambda},q)}\,q^{-\frac{D}{2}}\Bigr).
\end{align}
As a consequence of the shape invariance (\ref{AAd=AdA+E}), we obtain
for $n\geq s\geq 0$,
\begin{align}
  &V_s(z;\bm{\lambda},q)=q^{-s}V(z;q^{\frac{s}{2}}\bm{\lambda},q),\\
  &A_s(z;\bm{\lambda},q)=q^{-\frac{s}{2}}A(z;q^{\frac{s}{2}}\bm{\lambda},q),
  \quad
  A_s(z;\bm{\lambda},q)^{\dagger}
  =q^{-\frac{s}{2}}A(z;q^{\frac{s}{2}}\bm{\lambda},q)^{\dagger},\\
  &H_s(z;\bm{\lambda},q)
  =A_s(z;\bm{\lambda},q)^{\dagger}A_s(z;\bm{\lambda},q)
  +\mathcal{E}_s(\bm{\lambda},q)
  =q^{-s}H(z;q^{\frac{s}{2}}\bm{\lambda},q)+\mathcal{E}_s(\bm{\lambda},q),
  \label{Hs=qH+E}\\
  &\mathcal{E}_{s+1}(\bm{\lambda},q)
  =\mathcal{E}_s(\bm{\lambda},q)
  +q^{-s}\mathcal{E}_1(q^{\frac{s}{2}}\bm{\lambda},q),
  \label{E=E+E}\\
  &H_s(x;\bm{\lambda},q)\phi_{s,n}(z;\bm{\lambda},q)
  =\mathcal{E}_n(\bm{\lambda},q)\phi_{s,n}(z;\bm{\lambda},q),\\
  &A_s(z;\bm{\lambda},q)\phi_{s,s}(z;\bm{\lambda},q)=0,\\
  &A_s(x;\bm{\lambda},q)^{\dagger}\phi_{s+1,n}(z;\bm{\lambda},q)
  =(\mathcal{E}_n(\bm{\lambda},q)-\mathcal{E}_s(\bm{\lambda},q))
  \phi_{s,n}(z;\bm{\lambda},q).
  \label{Adphi}
\end{align}
The relation (\ref{E=E+E}) means that $\{\mathcal{E}_n\}$ is calculable 
from $\mathcal{E}_1$ (\ref{E1}).
In other words, the spectrum is determined by the shape invariance.

{}From (\ref{phi=Aphi}) and (\ref{Adphi}) we obtain formulas,
\begin{align}
  &\phi_{s,n}(z;\bm{\lambda},q)=A_{s-1}(z;\bm{\lambda},q)\cdots
  A_1(z;\bm{\lambda},q)A_0(z;\bm{\lambda},q)\phi_n(z;\bm{\lambda},q),
  \label{phi=AA..}\\
  &\phi_n(z;\bm{\lambda},q)=
  \frac{A_0(z;\bm{\lambda},q)^{\dagger}}
  {\mathcal{E}_n(\bm{\lambda},q)-\mathcal{E}_0(\bm{\lambda},q)}\,
  \frac{A_1(z;\bm{\lambda},q)^{\dagger}}
  {\mathcal{E}_n(\bm{\lambda},q)-\mathcal{E}_1(\bm{\lambda},q)}\cdots
  \frac{A_{n-1}(z;\bm{\lambda},q)^{\dagger}}
  {\mathcal{E}_n(\bm{\lambda},q)-\mathcal{E}_{n-1}(\bm{\lambda},q)}\,
  \phi_{n,n}(z;\bm{\lambda},q).
  \label{phi=AdAd..}
\end{align}
The former (\ref{phi=AA..}) gives the eigenfunction $\phi_{s,n}$ of the
$s$-th Hamiltonian $H_s$ along the isospectral line with energy
$\mathcal{E}_n$, starting from $\phi_n$ of the original Hamiltonian $H$
by repeated application of the $A$ operators.
The latter (\ref{phi=AdAd..}), on the other hand, expresses the $n$-th
eigenfunction $\phi_n$ of the original Hamiltonian, starting from the
explicitly known ground state $\phi_{n,n}$ of the $n$-th Hamiltonian
$H_n$ by repeated application of the $A^\dagger$ operators.
Since (\ref{Hs=qH+E}) implies $\phi_{n,n}(z;\bm{\lambda},q)\propto
\phi_0(z;q^{\frac{n}{2}}\bm{\lambda},q)$, $\phi_n$ is expressed in terms
of $\phi_0$ and $V$.
The latter formula (\ref{phi=AdAd..})
could also be understood as the generic form of the Rodrigue's formula
for the orthogonal polynomials.
The situation is depicted in Fig.1.
\begin{figure}
 \centering
 \includegraphics*[scale=.7]{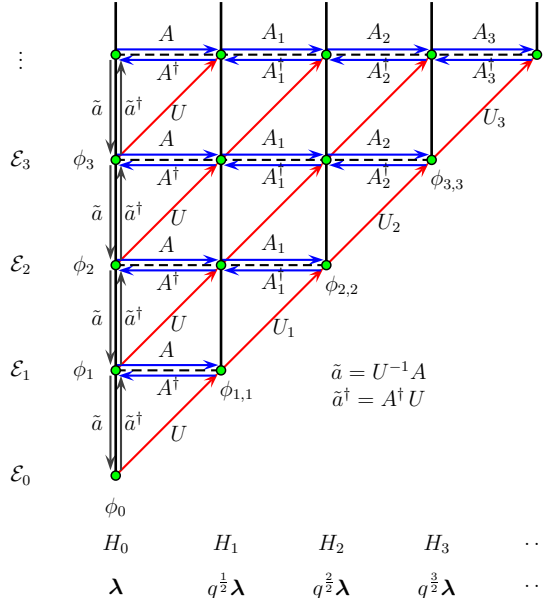}
 \caption{A schematic diagram of the energy levels and the associated
 Hamiltonian systems  together with the definition of the $A$ and
 $A^\dagger$ operators and the `creation' ($\tilde{a}^{\dagger}$) and
`annihilation' ($\tilde{a}$) operators.
 The parameter set is indicated below each Hamiltonian.}
 \label{fig:aadagdef}
\end{figure}
The operator $A$ acts to the right and $A^\dagger$ to the left along
the horizontal ({\em isospectral\/}) line.
They should not be confused with the {\em annihilation\/} and
{\em creation\/} operators, which act along the vertical line of
a given Hamiltonian $H_s$ going from one energy level $\mathcal{E}_n$
to another $\mathcal{E}_{n\pm 1}$.

In order to define the annihilation and creation operators, let us introduce
normalized basis $\{\hat{\phi}_{s,n}\}_{n\ge s}$ for each Hamiltonian
$H_s$. Ordinarily, the phase of each element of an orthonormal basis could
be completely arbitrary. In the present case, however, the eigenfunctions
are orthogonal polynomials. That is, they are real and the relations among
different degree members are governed by the three-term recurrence
relations. So the phases of $\{\hat{\phi}_{s,n}\}_{n\ge s}$ are fixed.
Let us introduce a unitary (in fact  an orthogonal) operator $U_s$ mapping
the $s$-th orthonormal basis $\{\hat{\phi}_{s,n}\}_{n\ge s}$ to the
$(s+1)$-th $\{\hat{\phi}_{s+1,n}\}_{n\ge s+1}$ (see Fig.~1 and for example
\cite{spivinzhed,kindao}):
\begin{equation}
  U_s\hat{\phi}_{s,n}=\hat{\phi}_{s+1,n+1},\quad
  U_s^\dagger \hat{\phi}_{s+1,n+1}=\hat{\phi}_{s,n}.
\end{equation}
We denote that $U_0=U$.
Roughly speaking $U$ changes the parameters from $\bm{\lambda}$ to
$q^{\frac12}\bm{\lambda}$.
Let us introduce an annihilation $\tilde{a}$ and a creation operator
$\tilde{a}^\dagger$ for the Hamiltonian $H$ as follows:
\begin{equation}
  \tilde{a}=\tilde{a}(z;\bm{\lambda},q)\eqdef U^{\dagger}A(z;\bm{\lambda},q),
  \quad
  \tilde{a}^{\dagger}=\tilde{a}(z;\bm{\lambda},q)^{\dagger}\eqdef
  A(z;\bm{\lambda},q)^{\dagger}U.
\end{equation}
It is straightforward to derive
\begin{align}
  &H=\tilde{a}^{\dagger}\tilde{a},\\
  &[\tilde{a},\tilde{a}^{\dagger}]\hat{\phi}_n(z;\bm{\lambda},q)
  =({\cal E}_{n+1}(\bm{\lambda},q)-{\cal E}_n(\bm{\lambda},q))
  \hat{\phi}_n(z;\bm{\lambda},q).
\end{align}

\section{Summary and Comments}

In this article we have studied the equilibrium positions of the
Ruijsenaars-Schneider-van Diejen systems with the trigonometric potential
and shown that for a suitable choice of the elementary potential
functions they are given by the zeros of the Askey-Wilson polynomials
with five parameters (see the footnote in section \ref{multiparticle}).
The equation for the equilibrium positions (\ref{equiveq}) (without
the positivity condition) can be written in the Bethe ansatz like equation
(\ref{equiveqBA}).
The Bethe ansatz is a powerful method for solvable models, and
solving the Bethe ansatz equation or clarifying the properties of its
solutions are very important.
Ismail et.al. studied Bethe ansatz equations for spin $s$ XXZ models
from the $q$-Sturm-Liouville problem point of view \cite{ismail}.
This kind of approach would shed new light on the Bethe ansatz(-like)
equations.

We have also studied the shape invariance of ``discrete" quantum
mechanical single particle systems with a $q$-shift type kinetic term.
As an example of this shape invariance, we present such a system whose
eigenfunctions are the Askey-Wilson polynomials.
In this example $V(z)$ is a rational function of $z$ (trigonometric
function of $\theta$), but the method works for a wider class of functions.
In ordinary quantum mechanics there is the Crum's theorem \cite{crum}, which
states a construction of the associated isospectral Hamiltonians $H_s$ and
their eigenfunctions $\phi_{s,n}$ (Fig.1) even if the system has no shape
invariance. The construction of $H_s$ and $\phi_{s,n}$ given in this
article and \cite{os4} needs shape invariance. A ``discrete" analogue of
the Crum's theorem, namely similar construction  without shape invariance,
would be very helpful, if exists.

We comment on the shape invariance of the Askey-Wilson polynomials with
a small number of parameters, $p_n(x;a,b,1,-1)$.
In this case $V(z)$ in (\ref{H}) is
\begin{equation}
  V(z;\bm{\lambda},q)=\frac{(1-az)(1-bz)}{1-qz^2},\quad\bm{\lambda}=(a,b).
\end{equation}
By taking a form (\ref{V1=Vg}) with the same $g(z)$ (\ref{g}),
the conditions (\ref{V1eq3}) and (\ref{V1eq4}) are satisfied, and we have
\begin{equation}
  V_1(z;\bm{\lambda},q)
  =q^{-1}V(q^{-\frac12}z;q\bm{\lambda},q),\quad
  \mathcal{E}_1(\bm{\lambda},q)=\frac12(q^{-1}-1)(1+ab).
\end{equation}
Since $D_z=z\frac{d}{dz}$ is invariant under the rescaling of $z$,
$D_z=D_{\alpha z}$, the Hamiltonian is shape invariant,
\begin{equation}
  H_1(z,\bm{\lambda},q)=q^{-1}H(q^{-\frac12}z;q\bm{\lambda},q)
  +\mathcal{E}_1(\bm{\lambda},q).
\end{equation}
The $s$-th Hamiltonian and the spectrum are given by
\begin{align}
  &H_s(z;\bm{\lambda},q)=q^{-s}H(q^{-\frac{s}{2}}z;q^s\bm{\lambda},q)
  +\mathcal{E}_s(\bm{\lambda},q),\\
  &\mathcal{E}_s(\bm{\lambda},q)
  =\mathcal{E}_{s-1}(\bm{\lambda},q)
  +q^{-(s-1)}\mathcal{E}_1(q^{s-1}\bm{\lambda},q)
  =\frac12q^{-s}(1-q^s)(1+abq^{s-1}).
\end{align}

\section*{Acknowledgements}

S. O. and R. S. are supported in part by Grant-in-Aid for Scientific
Research from the Ministry of Education, Culture, Sports, Science and
Technology, No.13135205 and No. 14540259, respectively.


\end{document}